\documentclass[fleqn,usenatbib]{mnras}
\usepackage{newtxtext,newtxmath}
\usepackage{multirow}
\usepackage[T1]{fontenc}
\DeclareRobustCommand{\VAN}[3]{#2}
\let\VANthebibliography\thebibliography
\def\thebibliography{\DeclareRobustCommand{\VAN}[3]{##3}\VANthebibliography}
\usepackage[dvipsnames]{xcolor}
\usepackage{graphicx}	
\usepackage{amsmath}	
\usepackage[referable]{threeparttablex}

\title[The hot corona in MCG-05-23-16]{The geometry of the hot corona in MCG-05-23-16 constrained by X-ray polarimetry}
\author[D. Tagliacozzo et al.]{
D. Tagliacozzo,$^{\ref{in:UniRoma3}}$  
A. Marinucci,$^{\ref{in:ASI}}$ 
F. Ursini,$^{\ref{in:UniRoma3}}$
G. Matt,$^{\ref{in:UniRoma3}}$  
S. Bianchi,$^{\ref{in:UniRoma3}}$  
L. Baldini,$^{\ref{in:INFN-PI}, \ref{in:UniPI}}$
T. Barnouin,$^{\ref{in:Strasbourg}}$ 
\newauthor
N. Cavero Rodriguez,$^{\ref{in:WUStL}}$ 
A. De Rosa,$^{\ref{in:INAF-IAPS}}$ 
L. Di Gesu,$^{\ref{in:ASI}}$ 
M. Dov\v{c}iak,$^{\ref{in:CAS-ASU}}$
D. Harper,$^{\ref{in:WUStL}}$ 
A. Ingram,$^{\ref{in:Newcastle}}$
V. Karas,$^{\ref{in:CAS-ASU}}$
\newauthor
D. E. Kim, $^{\ref{in:INAF-IAPS},\ref{in:SAPIENZA},\ref{in:UniRoma2}}$
H. Krawczynski ,$^{\ref{in:WUStL}}$
G. Madejski,$^{\ref{in:Stanford}}$
F. Marin,$^{\ref{in:Strasbourg}}$ 
R. Middei,$^{\ref{in:INAF-OAR},\ref{in:ASI-SSDC}}$
H. L. Marshall,$^{\ref{in:MIT}}$
F. Muleri,$^{\ref{in:INAF-IAPS}}$ 
\newauthor
C. Panagiotou,$^{\ref{in:MIT}}$
P.-O. Petrucci,$^{\ref{in:Grenoble}}$ 
J. Podgorny,$^{\ref{in:Strasbourg},\ref{in:CAS-ASU},\ref{in:Charles}}$ 
J. Poutanen,$^{\ref{in:UTU}}$
S. Puccetti,$^{\ref{in:ASI-SSDC}}$
P. Soffitta,$^{\ref{in:INAF-IAPS}}$ 
F. Tombesi,$^{\ref{in:UniRoma2},\ref{in:INFN-Roma2},\ref{in:UMd}}$
\newauthor
A. Veledina,$^{\ref{in:UTU},\ref{in:KTH}}$
W. Zhang,$^{\ref{in:NAO-CAS}}$
I. Agudo,$^{\ref{in:CSIC-IAA}}$
L. A. Antonelli,$^{\ref{in:INAF-OAR},\ref{in:ASI-SSDC}}$ 
M. Bachetti,$^{\ref{in:INAF-OAC}}$ 
W. H. Baumgartner,$^{\ref{in:NASA-MSFC}}$ 
\newauthor
R. Bellazzini,$^{\ref{in:INFN-PI}}$ 
S. D. Bongiorno,$^{\ref{in:NASA-MSFC}}$ 
R. Bonino,$^{\ref{in:INFN-TO},\ref{in:UniTO}}$
A. Brez,$^{\ref{in:INFN-PI}}$ 
N. Bucciantini,$^{\ref{in:INAF-Arcetri},\ref{in:UniFI},\ref{in:INFN-FI}}$ 
F. Capitanio,$^{\ref{in:INAF-IAPS}}$
\newauthor
S. Castellano,$^{\ref{in:INFN-PI}}$  
E. Cavazzuti,$^{\ref{in:ASI}}$ 
C.-T. Chen,$^{\ref{in:USRA-MSFC}}$
S. Ciprini,$^{\ref{in:INFN-Roma2},\ref{in:ASI-SSDC}}$
E. Costa,$^{\ref{in:INAF-IAPS}}$ 
E. Del Monte,$^{\ref{in:INAF-IAPS}}$ 
N. Di Lalla,$^{\ref{in:Stanford}}$
\newauthor
A. Di Marco,$^{\ref{in:INAF-IAPS}}$
I. Donnarumma,$^{\ref{in:ASI}}$
V. Doroshenko,$^{\ref{in:Tub}}$
S. R. Ehlert,$^{\ref{in:NASA-MSFC}}$  
T. Enoto,$^{\ref{in:RIKEN}}$
Y. Evangelista,$^{\ref{in:INAF-IAPS}}$
S. Fabiani,$^{\ref{in:INAF-IAPS}}$
\newauthor
R. Ferrazzoli,$^{\ref{in:INAF-IAPS}}$ 
J. A. Garcia,$^{\ref{in:Caltech}}$
S. Gunji,$^{\ref{in:Yamagata}}$  
J. Heyl,$^{\ref{in:UBC}}$
W. Iwakiri,$^{\ref{in:Chiba}}$ 
S. G. Jorstad,$^{\ref{in:BU},\ref{in:SPBU}}$ 
P. Kaaret,$^{\ref{in:NASA-MSFC}}$  
F. Kislat,$^{\ref{in:UNH}}$ 
\newauthor
T. Kitaguchi,$^{\ref{in:RIKEN}}$ 
J. J. Kolodziejczak,$^{\ref{in:NASA-MSFC}}$ 
F. La Monaca,$^{\ref{in:INAF-IAPS}}$ 
L. Latronico ,$^{\ref{in:INFN-TO}}$ 
I. Liodakis,$^{\ref{in:FINCA}}$
S. Maldera,$^{\ref{in:INFN-TO}}$  
\newauthor
A. Manfreda,$^{\ref{INFN-NA}}$
A. P. Marscher,$^{\ref{in:BU}}$ 
F. Massaro,$^{\ref{in:INFN-TO},\ref{in:UniTO}}$ 
I. Mitsuishi,$^{\ref{in:Nagoya}}$ 
T. Mizuno,$^{\ref{in:Hiroshima}}$ 
M. Negro,$^{\ref{in:UMBC},\ref{in:NASA-GSFC},\ref{in:CRESST}}$ 
\newauthor
C.-Y. Ng,$^{\ref{in:HKU}}$
S. L. O'Dell,$^{\ref{in:NASA-MSFC}}$  
N. Omodei,$^{\ref{in:Stanford}}$
C. Oppedisano,$^{\ref{in:INFN-TO}}$  
A. Papitto,$^{\ref{in:INAF-OAR}}$
G. G. Pavlov,$^{\ref{in:PSU}}$
A. L. Peirson,$^{\ref{in:Stanford}}$
\newauthor
M. Perri,$^{\ref{in:ASI-SSDC},\ref{in:INAF-OAR}}$
M. Pesce-Rollins,$^{\ref{in:INFN-PI}}$ 
M. Pilia,$^{\ref{in:INAF-OAC}}$ 
A. Possenti,$^{\ref{in:INAF-OAC}}$ 
B. D. Ramsey,$^{\ref{in:NASA-MSFC}}$  
J. Rankin,$^{\ref{in:INAF-IAPS}}$ 
\newauthor
A. Ratheesh,$^{\ref{in:INAF-IAPS}}$ 
O. J. Roberts,$^{\ref{in:USRA-MSFC}}$
R. W. Romani,$^{\ref{in:Stanford}}$
C. Sgr\`o,$^{\ref{in:INFN-PI}}$  
P. Slane,$^{\ref{in:CfA}}$  
G. Spandre,$^{\ref{in:INFN-PI}}$ 
\newauthor
D. A. Swartz,$^{\ref{in:USRA-MSFC}}$
T. Tamagawa,$^{\ref{in:RIKEN}}$
F. Tavecchio,$^{\ref{in:INAF-OAB}}$
R. Taverna,$^{\ref{in:UniPD}}$ 
Y. Tawara,$^{\ref{in:Nagoya}}$
A. F. Tennant,$^{\ref{in:NASA-MSFC}}$  
\newauthor
N. E. Thomas,$^{\ref{in:NASA-MSFC}}$  
A. Trois,$^{\ref{in:INAF-OAC}}$
S. S. Tsygankov,$^{\ref{in:UTU}}$
R. Turolla,$^{\ref{in:UniPD},\ref{in:MSSL}}$
J. Vink,$^{\ref{in:Amsterdam}}$
M. C. Weisskopf,$^{\ref{in:NASA-MSFC}}$ 
K. Wu,$^{\ref{in:MSSL}}$
\newauthor
F. Xie,$^{\ref{in:GSU},\ref{in:INAF-IAPS}}$
S. Zane$^{\ref{in:MSSL}}$
}

\date{Accepted XXX. Received YYY; in original form ZZZ}

\pubyear{2023}

\begin{document}
\label{firstpage}
\pagerange{\pageref{firstpage}--\pageref{lastpage}}
\maketitle

\begin{abstract}
We report on the second observation of the radio-quiet active galactic nucleus (AGN) MCG-05-23-16 performed with the \textit{Imaging X-ray Polarimetry Explorer} (\textit{IXPE}). The observation started on 2022 November 6 for a net observing time of 640 ks, and was partly simultaneous with \textit{NuSTAR} (86 ks). After combining these data with those obtained in the first  \textit{IXPE} pointing on May 2022 (simultaneous with \textit{XMM-Newton} and \textit{NuSTAR}) we find a 2--8 keV polarization degree $\Pi=1.6\pm0.7$ (at 68 per cent confidence level), which corresponds to an upper limit $\Pi=3.2$ per cent (at 99 per cent confidence level). We then compare the polarization results with Monte Carlo simulations obtained with the \textsc{monk} code, with which different coronal geometries have been explored (spherical lamppost, conical, slab and wedge). Furthermore, the allowed range of inclination angles is found for each geometry. If the best fit inclination value from a spectroscopic analysis is considered, a cone-shaped corona along the disc axis is disfavoured.
\end{abstract}

\begin{keywords}
galaxies: active -- galaxies: Seyfert -- polarization -- X-rays:galaxies -- X-rays: individual: MCG-05-23-16
\end{keywords}

\section{Introduction}

The large amount of energy released by AGNs is widely thought to be generated in a very compact and central region via accretion onto a supermassive black hole (SMBH, \citealt{Rees84, 1993ARA&A..31..473A}). The optical/UV radiation emitted by the accretion disc is partly redirected towards the X-ray band  (primary emission) through a process known as Comptonization, which involves multiple scatterings in a cloud of hot electrons, generally called the corona \citep{1980A&A....86..121S, 1991ApJ...380L..51H, 2000ApJ...542..703Z, zdz10.1143/PTPS.155.99,Done_2007}. These structures are characterized by high electron temperatures ($kT_{\rm e}$ usually ranging from tens to hundreds keV) and moderate Thomson optical depths ($\tau$, \citealt{Petrucci_2001, perolarefId0, dadarefId0, pan2011MNRAS.417.2426P, dero2012MNRAS.420.2087D, ricci2017ApJS..233...17R, galaxies6020044, refId0, middei2019A&A...630A.131M}). Despite being a key element in understanding the energy generation mechanism of AGNs, the morphology of the corona, which may hold clues to its physical origins, remains a matter of debate. While in principle spectroscopic techniques can provide information on the coronal geometry, even
the best observations, while providing valuable information on its physical parameters such as temperature and optical depth, fall short of distinguishing
between different geometrical configurations \citep{2019ApJ...875..148Z, refId0}. Currently, some constraints on the coronal morphology have been derived using time lags techniques (such as reverberation mapping, \citealt{uttarticle, fa2017AN....338..269F, cab2020MNRAS.498.3184C}), but many aspects remain to be determined. In this context, X-ray polarimetry represents a fundamental tool in order to investigate the coronal properties and constrain its geometry, because different morphologies of the emitting region produce different polarization signatures.

 Several geometrical models have been proposed for the corona. In this work we consider the following: spherical lamppost, conical outflow, slab corona and wedge-shaped hot accretion flow. The spherical lamppost consists of an isotropic spherical source located on the spin axis of the SMBH \citep{sph1991A&A...247...25M, 2012MNRAS.424.1284W, 2022MNRAS.510.3674U} and it is defined by its radius and its height above the SMBH. 
 This configuration is expected to produce a low polarization degree ({PD}$=0-2$ per cent) with the polarization angle (PA) perpendicular to the accretion disc axis \citep{2022MNRAS.510.3674U}. The conical outflow is commonly associated with an aborted jet \citep{1997A&A...326...87H, conerefId0, 2022MNRAS.510.3674U}. According to this model, radio-quiet AGNs have central SMBHs powering outflows and jets which may propagate only for a short distance, if the velocity of the ejected material is sub-relativistic and smaller than the escape velocity. 
This configuration is expected to produce somewhat larger (up to 6 per cent) polarization degree, also in this case perpendicular to the accretion disc axis  \citep{2022MNRAS.510.3674U}. In the slab corona scenario the hot medium is assumed to be uniformly distributed above the cold accretion disc. This geometry can be realised in the scenario where magnetic loops rise high above the disc plane and dissipate energy via reconnection \citep{1979ApJ...231L.111L, 1991ApJ...380L..51H, belo2017ApJ...850..141B}. This configuration can produce polarization degree up to 14 per cent  \citep{sve1996ApJ...470..249P, 2022MNRAS.510.3674U, gia2023arXiv230312541G}. In this case the polarization angle is parallel to the accretion disc axis. The wedge is, finally, similar to the slab but with the height increasing with the radius. In this scenario the `standard' accretion disc is thought to be truncated at a certain radius, while the corona represents some type of a `hot accretion flow', possibly extending to the innermost stable circular orbit (ISCO, \citealt{Esin_1997, sh2010ApJ...712..908S, krol1997MNRAS.292L..21P, Yuan2014, poutanen2018, popurefId0}). It is expected to produce intermediate (up to 5 per cent, depending on the specific assumed configuration) polarization degree, parallel to the accretion disc axis. This configuration is considered in detail in Sect. \ref{sim}. 

MCG-05-23-16 is a nearby ($z=0.0085$, \citealt{2003AJ....126.2268W}) Seyfert 1.9 galaxy \citep{1980A&A....87..245V} with broad emission lines in the infrared \citep{infr1994ApJ...422..521G}. It is a relatively bright X-ray source ($F_{2-10}=7-10\times10^{-11}$erg cm$^{-2}$s$^{-1}$, \citealt{2004ApJ...601..771M}) showing moderate cold absorption ($N_{\rm H}\sim10^{22}\,\mbox{cm}^{-2}$). It has been widely studied in the X-ray band \citep{be2008A&A...492...93B,moli2013MNRAS.433.1687M}, and its high energy cut-off ($E_{\rm C}$) and coronal physical parameters, i.e. temperature and Thomson optical depth, are quite well estimated (\citealt{2015ApJ...800...62B}). The SMBH mass ($M_{\rm BH}=2\times10^7\,\mbox{M}_{\odot}$) has been estimated via X-ray variability \citep{po2012A&A...542A..83P}, and it is consistent with the virial mass derived from the infrared lines \citep{ono2017MNRAS.468L..97O}. From the observations of MCG-05-23-16 performed with \textit{XMM-Newton}, \textit{NuSTAR} and \textit{IXPE} in May 2022, \citet{Marinucci_2022} found, assuming a simple cut-off power law for the primary continuum, a spectral index $\Gamma=1.85\pm0.01$ and a high energy cut-off $E_{\rm C}=120\pm15$ keV, leading to an electron temperature $kT_{\rm e}=25\pm2$ keV and $\tau=1.27\pm0.08$ if the cut-off power law is replaced by the comptonization model \textsc{compps} \citep{sve1996ApJ...470..249P} and a uniform slab geometry for the corona is assumed. Moreover, a 4.7  per cent upper limit (99  per cent c.l. for one parameter of interest) for the polarization degree was obtained.

In this paper we present and discuss the second \textit{IXPE} observation of MCG-5-23-16, performed in November 2022 in coordination with \textit{NuSTAR}. 
The combined analysis of the data collected in the 2022 May and November  observations are also discussed. The results are then compared with Monte Carlo simulations of the expected polarization properties for different geometries of the corona. 

The paper is organized as follows: in Sect. \ref{obs} we discuss the data reduction procedure, in Sect. \ref{analisi} we present the spectropolarimetric data analysis, in Sect. \ref{sim} we present
Monte Carlo simulations designed to calculate the expected polarization for different geometries and, finally, the results are  summarized in Sect. \ref{discussion}.

\section{OBSERVATIONS AND DATA REDUCTION}
\label{obs}
\begin{figure*}
\includegraphics[width=1.95\columnwidth] {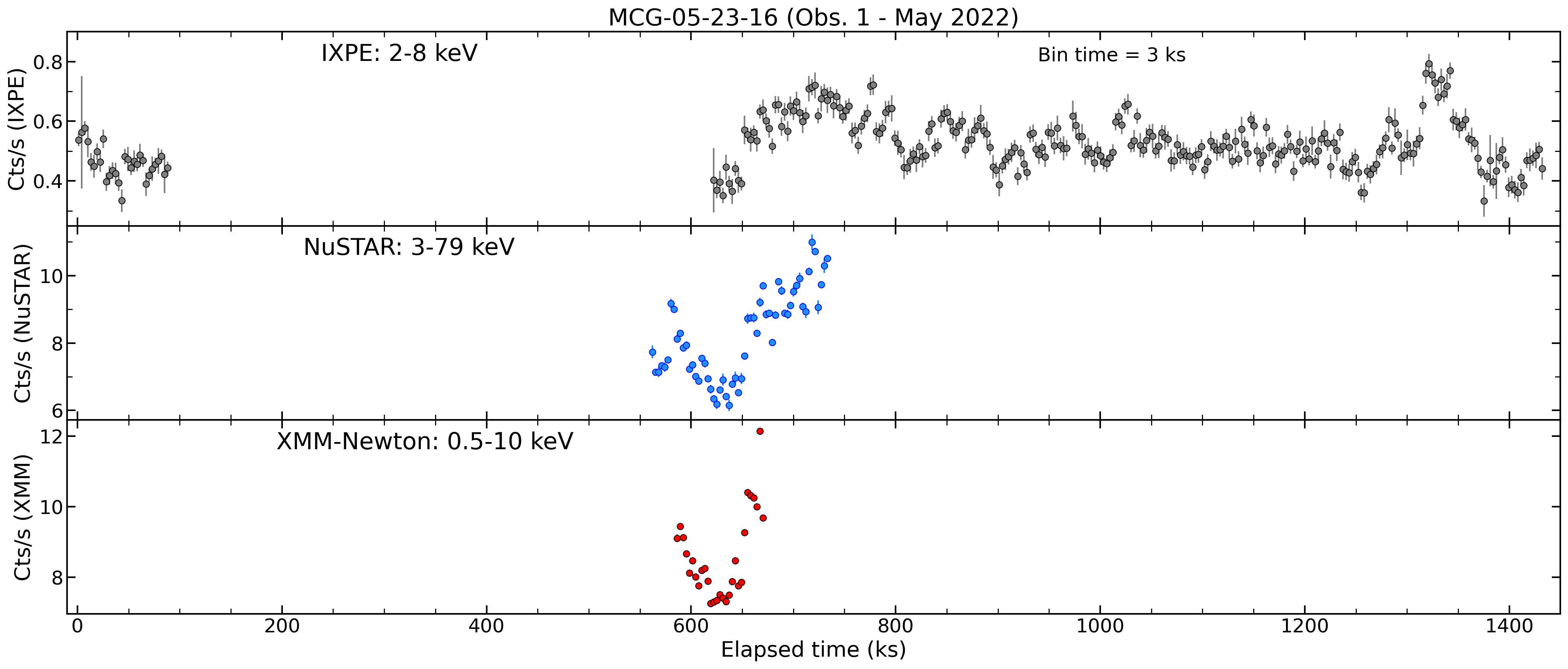}
\includegraphics[width=1.95\columnwidth] {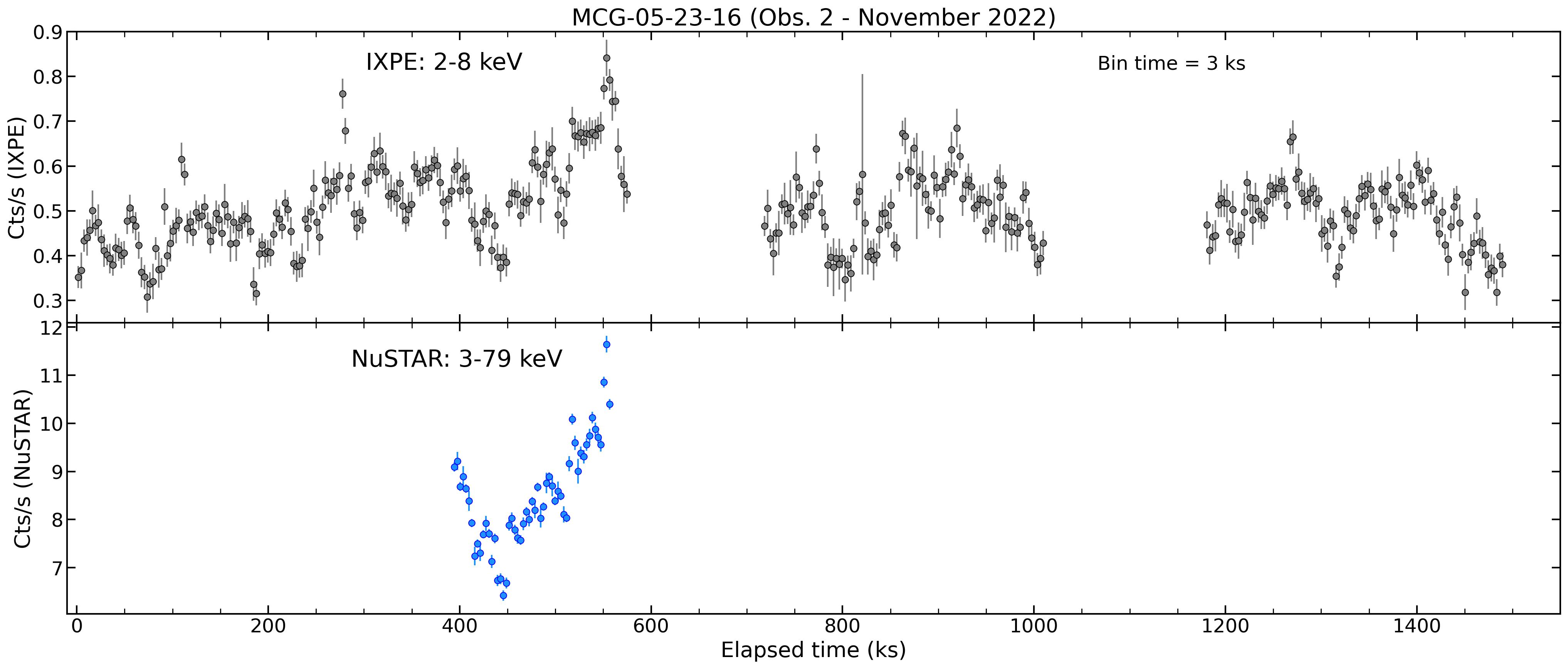}
\caption{\textit{IXPE}, {\it NuSTAR} and \textit{XMM-Newton} light curves of the two observing campaign of MCG-05-23-16 are shown. Data counts from DU1, DU2 and DU3 on board on \textit{IXPE} and from FPMA/B on board on {\it NuSTAR} have been summed. The full energy bands of the three satellites have been used and we adopted a 3 ks time binning.}\label{lc}
\end{figure*}

{\it IXPE} \citep{ixpe} observed MCG-05-23-16 twice, in May and November 2022. The first {\it IXPE} observation and the simultaneous \textit{XMM-Newton} and {\it NusTAR} data are presented in \cite{Marinucci_2022}. These spectra, with updated response matrices, are also used in this work. The second pointing started on November 6, and had a net exposure time of 642 ks. 
Cleaned level 2 event files were produced and calibrated using standard filtering criteria with the dedicated {\sc ftools} tasks and the latest calibration files available in the {\it IXPE} calibration database (CALDB 20220303). $I$, $Q$ and $U$ Stokes background spectra were extracted from source-free circular regions with a radius of 100 arcSect. Extraction radii for the $I$ Stokes spectra of the source were computed via an iterative process which leads to the maximization of the Signal-to-Noise Ratio (SNR) in the 2--8 keV energy band, similar to the approach described in \citet{pico04}. We therefore adopted  circular regions centered on the source with radii of 62 arcsec for the three DUs. The net exposure times are 641.7 ks and the same extraction radii were then applied to the $Q$ and $U$ Stokes spectra. We used a constant energy binning of 0.2 keV for $Q$, $U$ Stokes spectra and required a SNR higher than 5 in each spectral channel, in the intensity spectra. $I$, $Q$, $U$ Stokes spectra from the three DUs are always fitted independently in the following, but we will plot them together using the {\sc setp group} command in {\sc xspec}, for the sake of visual clarity. Background represents
2.0, 1.8 and 2.1  per cent of the total DU1, DU2 and DU3 $I$ spectra, respectively. We followed the formalism discussed in \citet{stro17} and used the weighted analysis method presented in \citet{dimarco22} (parameter {\sc stokes=Neff} in {\sc xselect}). The summed background subtracted light curves for the two {\it IXPE} poitings are shown in Fig. \ref{lc}.

{\it NuSTAR} \citep{nustar} observed MCG-05-23-16, with its two coaligned X-ray telescopes with corresponding Focal Plane Module A (FPMA) and B (FPMB), on 2022 November 11. The total elapsed time is 164.6 ks. The Level 1 data products were processed with the {\it NuSTAR} Data Analysis Software (NuSTARDAS) package (v. 2.1.2). Cleaned event files (level 2 data products) were produced and calibrated using standard filtering criteria with the \textsc{nupipeline} task and the latest calibration files available in the {\it NuSTAR} calibration database (CALDB 20221020). Extraction radii for the source and background spectra were $40$ arcsec and 60 arcsec, FPMA spectra were binned in order not to over-sample the instrumental resolution more than a factor of 2.5 and to have a SNR  greater than 5 in each spectral channel, the same energy binning was then applied to the FPMB spectra. The net observing times for the FPMA and the FPMB data sets are 85.7 ks and 84.9 ks, respectively. The summed background subtracted FPMA and FPMB light curves are shown in Fig. \ref{lc}. We adopt the cosmological parameters $H_0=70$ km s$^{-1}$ Mpc$^{-1}$, $\Omega_\Lambda=0.73$ and $\Omega_{\rm m}=0.27$, i.e. the default ones in \textsc{xspec 12.12.1} \citep{1996ASPC..101...17A}. Errors correspond to the 90  per cent confidence level for one interesting parameter ($\Delta\chi^2=2.7$), if not stated otherwise. 
\section{DATA ANALYSIS}
\label{analisi}

\subsection{\textit{IXPE} analysis}
\begin{figure}
\includegraphics[width=8.8cm] {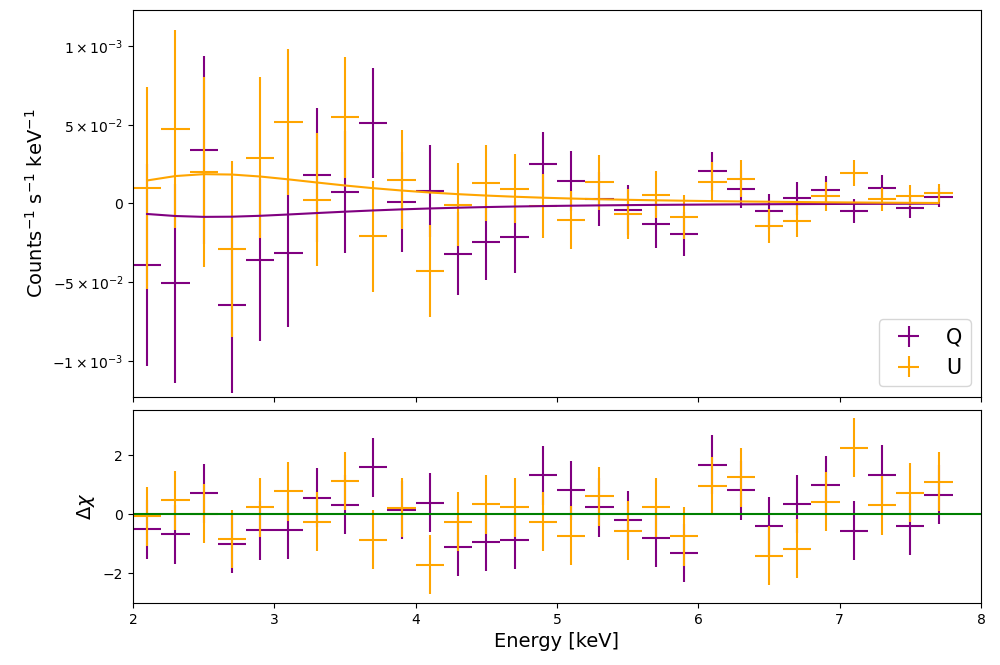}
\caption{\textit{IXPE} $Q$ (purple crosses) and $U$ (orange crosses) grouped Stokes spectra of the second IXPE pointing (November 2022) of MCG-05-23-16 are shown with residuals, along with the corresponding best-fitting model.}\label{stokes_par}
\end{figure}

\begin{figure}
\includegraphics[width=8.8cm] {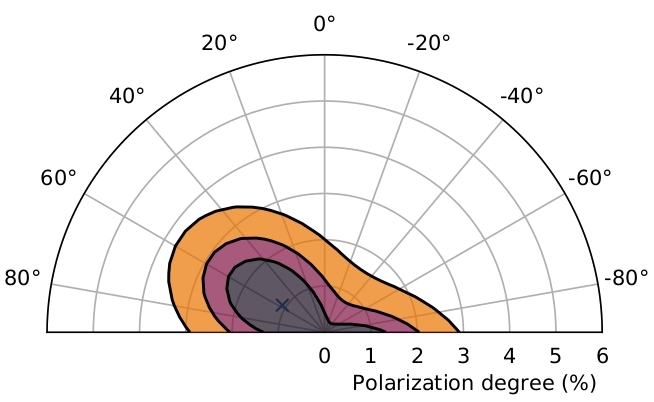}
\caption{Contour plot between the polarization degree $\Pi$ and angle $\Psi$ for the November 2022 data. Purple, pink and orange regions correspond, respectively, to 68, 90 and 99  per cent confidence levels for two parameters of interest.}\label{ixpe_nov_combined}
\end{figure}

Initially, we conducted a preliminary examination of the \textit{IXPE} data through the utilization of a baseline model, consisting of an absorbed power law convolved with a constant polarization kernel: $\textsc{const} \times \textsc{polconst} \times \textsc{tbabs} \times \textsc{powerlaw}$. We fit this model in the $2-8$ keV energy range simultaneously to the $I$, $Q$ and $U$ spectra collected by the 3 \textit{IXPE} Detector Units (DUs) during the second observation of MCG-05-23-16 (2022 November 6; 640ks). In all cases where we only use \textit{IXPE} data the adoption of a more complex model is unnecessary.
In fact, the reduced chi-square is always close to unity.
At the 68  per cent c.l for one parameter of interest, we obtained a polarization degree $\Pi=1.1\pm0.9$  per cent and a polarization angle $\Psi=57\degr\pm27\degr$. This translates into a 99  per cent c.l. upper limit to the polarization degree of $\Pi=3.3$  per cent. In Fig. \ref{stokes_par} we show the $Q$ and $U$ spectra used to perform this first analysis (along with the model and the residuals), while in Fig. \ref{ixpe_nov_combined} the contour plot between $\Pi$ and $\Psi$ is shown. An alternative, model-independent, analysis of the polarization cubes with the software {\sc ixpeobssim} \citep{baldini22} gives consistent results.

We then performed a combined analysis of the \textit{IXPE} $I$, $Q$ and $U$ spectra collected on May and November 2022, using the same model.
We notice a significant variation of the primary continuum spectral index between the two pointings. This variation is only found in the \textit{IXPE} data. For this reason, we consider it as a calibration issue concerning the first pointing of MCG-05-23-16 by this instrument. For this reason, it is not possible to sum the two observations together. We therefore proceeded to conduct a combined analysis leaving untied the spectral indexes
of the two pointings. We obtained  (at 68  per cent c.l. for one interesting parameter) a polarization degree $\Pi=1.6\pm0.7$  per cent and a polarization angle $\Psi=53\degr\pm13\degr$. This translate into a polarization degree upper limit (at 99  per cent c.l.) $\Pi=3.2$ per cent. This represents a significant improvement with respect to the results obtained for the May 2022 observation alone.
In Table \ref{pol_ixpe} the best-fit values of the polarization degree and angle obtained using only \textit{IXPE} dataset are shown.

\medskip

\begin{table}
\centering
\caption{Polarimetric properties of MCG-05-23-16 obtained with \textit{IXPE}. }
\begin{tabular}{lccc}
\hline
{Parameter} & {May 2022} & {Nov 2022} & {May+Nov 2022} \\ [0.5ex]
\hline
$\Pi$ (\%) & $2.2\pm1.7$& $1.1\pm0.9$ & $1.6\pm0.7$ \\ [1ex]
$\Psi$  (deg)  & $50\pm24$ &$57\pm27$ & $53\pm13$ \\ [1ex]
$\Pi$ (\%)  & $\le4.7$ &$\le3.3$ & $\le3.2$ \\ [1ex]
\hline
\end{tabular}
\begin{flushleft}{
\textit{Note}: The errors are shown at 68 per cent and the upper limits at 99 per cent confidence level for one parameter of interest.
}\end{flushleft}  
\label{pol_ixpe}
\end{table}

\subsection{\textit{XMM-Newton}, \textit{NuSTAR} and \textit{IXPE} combined analysis}
\label{combined_analysis}

As a next step we performed a spectropolarimetric analysis combining the 2--8 keV \textit{IXPE} spectra (May+November), the 2--10 keV \textit{XMM-Newton} spectrum (May) and the 3--79 keV \textit{NuSTAR} spectra (May+November). 
Taking advantage of the previous analysis of the May observations \citep{Marinucci_2022}, we used the following model: 

\noindent $\textsc{const}\times\textsc{tbabs}[\textsc{polconst}\times\textsc{ztbabs}\times\textsc{cutoffpl}+$

\noindent $\textsc{vashift}(\textsc{polconst}\times\textsc{kerrdisk}+\textsc{polconst}\times\textsc{xillver})]$,

\noindent where the constant component is needed to cross-calibrate the data set collected by the different detectors (DU1, DU2, DU3, FPMA, FPMB and EPIC pn). The primary continuum is modeled using a simple power law with a high energy exponential cut off (\textsc{cutoffpl}), while \textsc{tbabs} is used to model the Galactic absorption, using a column density $N_{\rm H}=7.8\times10^{20}\,\mbox{cm}^{-2}$ \citep{nh2016A&A...594A.116H}. The reflection from distant (and neutral) material (such as the external regions of the accretion disc and the torus) is modeled using \textsc{xillver} \citep{xill2013ApJ...768..146G}. The spectral index and high energy cut off in the reflection model are linked to those of the primary emission. The Fe abundance is set equal to the solar value and the inclination angle to $\theta=30\degr$. The \textsc{kerrdisk} component \citep{kerr2006ApJ...652.1028B} is used to deal with some residuals close to 6.4 keV, which may be interpreted as a Fe K$\alpha$ line from the inner part of the accretion disc, broadened by relativistic effects. For the \textit{XMM-Newton} spectrum we added a \textsc{vashift} component (which simply provides a shift in energy) in order to deal with the energy of the narrow Fe K$\alpha$ line, which is inconsistent with being 6.4 keV in the host galaxy rest frame. This effect is only found in the pn (and not in the MOS), so we conclude that it is likely due to calibration issues. We noticed a similar effect also in \textit{NuSTAR}, with an increasing deviation between the first and the second pointing. For this reason, we added a \textsc{vashift} component here too, attributing the effect to instrument degradation in time. In \textsc{kerrdisk}, the black hole spin is fixed to $a=0.998$, since the fit is largely insensitive to this parameter. Moreover, we fixed the disc emissivity profile to $\epsilon(r)\propto r^{-3}$. The rest frame energy of the line was fixed to 6.4 keV and the inner radius of the disc to its previously found best-fit value ($37 R_{\rm G}$, as found by \citealt{re2007PASJ...59S.301R}) \footnote{A complete and detailed spectroscopic analysis of these datasets, including the relativistic effects, will be presented in a forthcoming paper (Serafinelli et al, in prep.)}. In order to deal with calibration issues that affect the spectral index of May \textit{IXPE} observation, we modified, as in \cite{Marinucci_2022}, the response files gain in the $I$ spectra (using the \textsc{gain fit} command). Finally, as done for the \textit{IXPE} analysis, we untied the primary continuum spectral indices between the two observations.

\begin{figure}
\includegraphics[width=9cm,height=7.2cm] {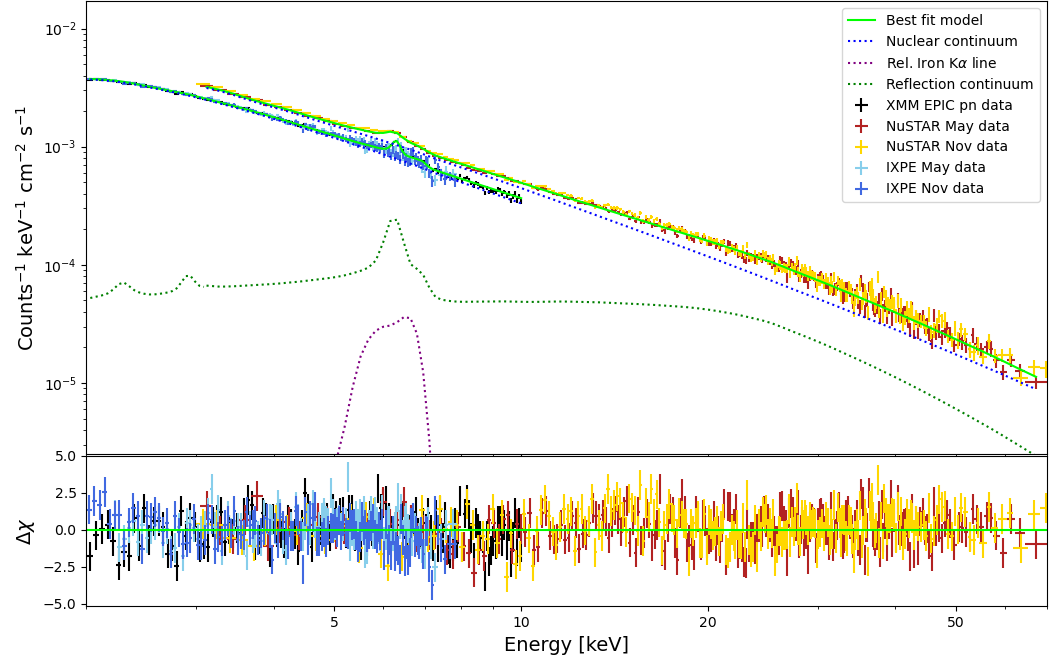}
\caption{The EPIC pn (May 2022), \textit{NuSTAR} (May+November 2022) and \textit{IXPE} $I$ (May+November 2022) spectra together with the best-fitting model  (\textit{upper panel}),  and the residuals (\textit{lower panel}).}\label{data_del}
\end{figure}

Each main spectral component (i.e. primary continuum and reflection) is associated with a different polarization. The Fe K$\alpha$ line is expected to be unpolarized \citep{Goosmann2011, marin2018A&A...615A.171M}, while the Compton reflection continuum contributes little in the {\it IXPE} band pass \citep{mar2018MNRAS.478..950M}. For these reasons, after checking the insensitivity of the fit to variations of these parameters, we fix the polarization of \textsc{kerrdisk} and \textsc{xillver} to zero for simplicity (see also \citealt{Marinucci_2022}). We get 
only an upper limit (at 99  per cent c.l. for one interesting parameter) for the polarization degree of the primary continuum of $\Pi=3.3$  per cent. At 68  per cent of c.l., we retrieve a polarization degree and angle of $\Pi=1.6\pm0.7$  per cent and $\Psi=53\degr\pm12\degr$, respectively. The fit is not ideal
($\chi ^2/$dof$=2381/2259$ (see Fig. \ref{data_del}) 
 but, since there is no evidence from the residuals of missing or wrong components in the model, we attribute it to an imperfect cross calibration between the three instruments.

\begin{figure}
\includegraphics[width=8.8cm] {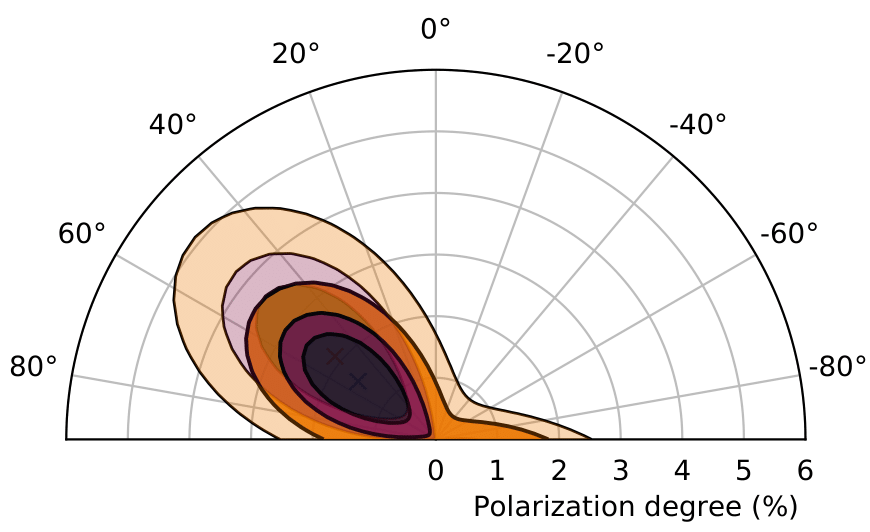}
\caption{Comparison between the polarization degree $\Pi$ and angle $\Psi$ contour plots from the combined (May+November 2022) observations  \textit{XMM-Newton}, \textit{NuSTAR} and \textit{IXPE} (\textit{saturated plot}) and the first (May 2022) observation only (\textit{pale plot}). Purple, pink and orange regions represent, respectively, the 68, 90 and 99 per cent confidence levels for two parameters of interest.}\label{fine_mondo}
\end{figure}

In Table \ref{new_data} we summarize the best-fitting values for all the free parameters of
this last and complete analysis (with errors at 68  per cent c.l.).
In Fig. \ref{fine_mondo} we show the contour plot of the polarization degree and angle of the continuum component, as well as a comparison with the contour plot from the May observation alone (\citet{Marinucci_2022}).

\begin{table}
\centering
\caption{Best-fitting parameters for the \textit{XMM-Newton}, \textit{NuSTAR} and \textit{IXPE} May+November 2022 combined  data set.}
\begin{tabular}{cc}
\hline
{Parameter}  & {Best fitting value} \\
\hline
$N_{\rm H}$ [cm$^{-2}$] & ($1.30\pm0.02$)$\times10^{22}$  \\
\hline
 $\Gamma_{\textsc{cutoffpl}}$ (May)& $1.84\pm0.01$ \\
 $\Gamma_{\textsc{cutoffpl}}$ (Nov)& $1.85\pm0.01$ \\
 \hline
 $E_{\rm C}$ [keV]& $120^{+9}_{-5}$  \\
 \hline
 $\Pi_{\textsc{cutoffpl}}$ [$\%$]&$1.6\pm0.7$  \\
 $\Psi_{\textsc{cutoffpl}}$ [deg]& $53\pm12$ \\
  $\Pi_{\textsc{\textsc{xill}}}=\Pi_{\textsc{\textsc{kerr}}}$ [$\%$]&0  \\
 $\Psi_{\textsc{xill}}=\Psi_{\textsc{kerr}}$ [deg]& 0 \\
 \hline
 \multicolumn{2}{c}{$v_{\textnormal{shift}}$ [km s$^{-1}$]} \\ [0.4ex]
 \medskip
 \textit{XMM-Newton} & $2.2^{+0.3}_{-0.4}\times 10^3$  \\
 \medskip
\textit{NuSTAR} (May) & $3.4^{+0.9}_{-0.4}\times10^3$  \\
\textit{NuSTAR} (Nov) & $5.5^{+1.0}_{-0.8}\times10^3$  \\
  \hline
  $\theta_{\textsc{kerr}}$ [deg]&$61^{+4}_{-13}$\\
  $a$&0.998\\
  $R_{\textnormal{in}}$ [$R_{\rm G}$]&$37$\\
  $\theta_{\textnormal{incl}}$ [deg]& 30\\
  \hline
   \multicolumn{2}{c}{NORMALIZATION CONSTANTS} \\ [0.4ex]
 $N_{\textsc{cutoffpl}}$&($2.52\pm0.02$)$\times10^{-2}$  \\
  $N_{\textsc{xill}}$&($2.0\pm0.1$)$\times10^{-4}$   \\
  $N_{\textsc{kerr}}$&($3.6\pm0.3$)$\times10^{-5}$    \\
  \hline
   \multicolumn{2}{c}{ $F_{2-10}$ [erg cm $^{-2}$s$^{-1}$]} \\ [0.4ex]
  \textit{XMM-Newton}& $(7.48\pm0.01)\times10^{-11}$\\
  \textit{NuSTAR} (Nov) & $(1.12\pm0.02)\times10^{-10}$\\
  \hline
  $L_{2-10}$ [erg s$^{-1}$]& $(1.70\pm0.01)\times10^{43}$\\
   $R$& $0.42\pm0.03$\\
  \hline
  $\chi^{2}$/dof& 2381/2259\\
  \hline
\end{tabular}
\begin{flushleft}{\textit{Note}: The errors at 68 per cent c.l. for one parameter of interest. $\Pi$ and $\Psi$ of \textsc{xillver} and \textsc{kerrdisk} are set equal to 0. Parameters without error have been frozen in the fit. The spectral index for the first observation is obtained applying the \textsc{gain fit} command. $R$ is the reflection fraction defined as the ratio between the 20--40 keV fluxes of the Compton reflection and the primary component.
}\end{flushleft} 
\label{new_data}
\end{table}

\section{Monte Carlo Simulations}
\label{sim}

\begin{figure*}
\includegraphics[width=15cm] {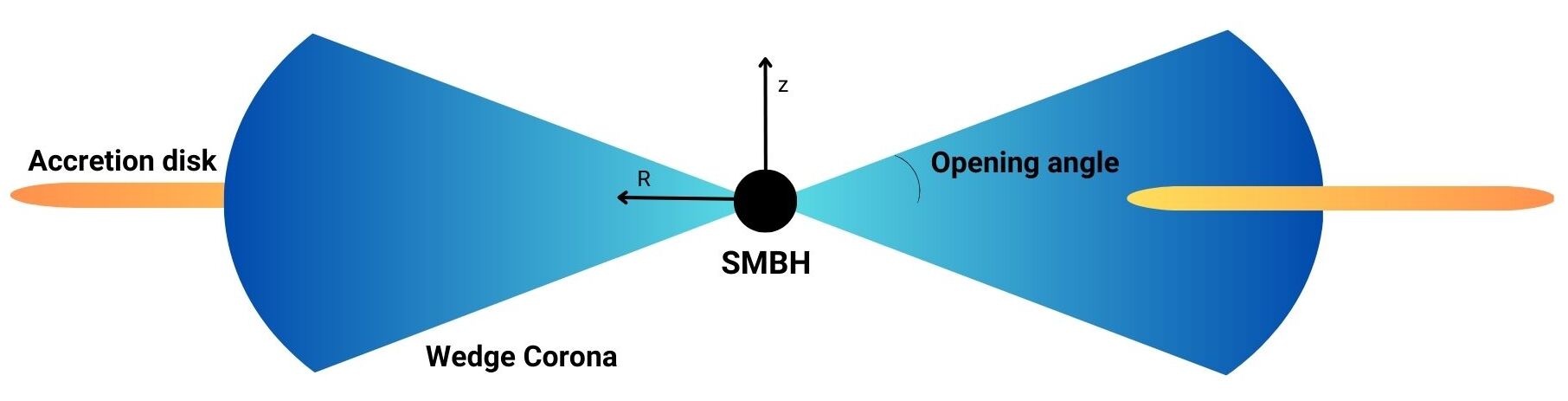}
\caption{The wedge corona. This geometry is characterized by an inner and an outer radius and an opening angle (measured from the accretion disc plane). In the left configuration, the inner radius of the accretion disc coincides with the outer radius of the corona, while in the right configuration it extends into the corona itself. The $R$ and $z$ axes represent the radial and the vertical coordinates.}\label{wedge_pic}
\end{figure*}

To interpret the polarization results, we perform detailed numerical simulations with the Monte Carlo code \textsc{monk} \citep{2019ApJ...875..148Z}, following the approach of \cite{2022MNRAS.510.3674U} (where spherical lamppost, conical outflow and slab have been already explored). We focus here on the so-called concave wedge geometry which, similarly to the slab, gives rise to polarization angles parallel to the accretion disc axis. A wedge configuration could potentially solve some of the theoretical issues that arise when using geometries such as the slab or the sphere \citep{Stern_1995, Done_2007, poutanen2018}.  The wedge geometry is defined by three parameters: an inner radius ($R_{\textnormal{in}}$), an outer radius ($R_{\textnormal{out}}$), and an opening angle ($\alpha$) (see Fig. \ref{wedge_pic}). We assume the inner radius to coincide with the Innermost Stable Circular Orbit (ISCO), which depends on the SMBH spin value (6 $R_{\rm G}$ for $a=0$ and 1.24 $R_{\rm G}$ for $a=0.998$). Unlike the slab configuration, the height of the wedge increases with radius. In this configuration the accretion disc is assumed to be truncated at a certain radius, while the corona represents a 'hot accretion flow', extending to the ISCO. The density profile of the wedge corona is uniform and the Thomson optical depth is computed radially. Finally, the accretion disc truncation radius can either coincide with the external edge of the corona or reach lower values, down to the ISCO. In Figure \ref{wedge_pic} a sketch of the wedge corona is shown. 

We perform Monte Carlo simulations for a total of 8 parameter combinations, considering only the external disc scenario with uniform coronal density (a detailed analysis of the various wedge configurations is beyond the scope of this paper and will be presented in a following paper). The simulations are run for two values of the SMBH spin ($a=0$ and $a=0.998$). In both cases, we set the inner radius to the ISCO, i.e. 6 $R_{\rm G}$ for the static black hole and 1.24 $R_{\rm G}$ for the maximally rotating black hole. We test four different opening angles ($15\degr$, $30\degr$, $45\degr$ and $60\degr$). We set the coronal electron temperature to 25 keV, as measured by \cite{2015ApJ...800...62B} and \cite{Marinucci_2022}.
After setting the electron temperature, for each geometrical configuration we find the optical depth that fits the spectrum best in the \textit{IXPE} band pass (i.e. 2--8 keV) when we replace the cut-off power law with the spectra obtained with \textsc{monk} in the best-fit model retrieved in Sect. \ref{combined_analysis}. In Table \ref{monk_par} we summarize the physical and geometrical parameters we assume in the simulations. For all the simulations we perform, we assume a mass of the SMBH of $M_{\rm BH}=2\times10^7M_{\odot}$ and an Eddington ratio of 0.1 \citep{po2012A&A...542A..83P}. Finally, we set the initial polarization (i.e. the polarization of the optical/UV radiation emitted by the accretion disc) as appropriate for a pure scattering, plane-parallel, semi infinite atmosphere \citep{1960ratr.book.....C}.

\begin{table}
\caption{Coronal input parameters for the \textsc{monk} simulations. } 
\label{monk_par}
\begin{tabular}{cccccc}
\hline
                 $kT_{\rm e}$ [keV] &    SMBH spin & $R_{\rm in}$ [$R_{\rm G}$] & $\alpha$ [deg] & $\tau$ & PD$_{\rm max}$ [\%] \\
\hline
\multirow{8}{*}{25} & \multirow{4}{*}{0} & \multirow{4}{*}{6}  &  15& 6.8 & 5 \\
                  &                  &                  & 30 &  4.2 & 4\\
                   &                  &                  & 45 &  3.3 & 3\\
                  &                   &                   & 60 & 2.8& 2.3\\

                  & \multirow{4}{*}{0.998} & \multirow{4}{*}{1.24} & 15  & 8.3 & 5.8 \\
                  &                   &                   &  30&5.1& 4.3 \\
                   &                   &                   &  45&3.8& 3.2 \\
                  &                   &                   & 60 & 3.2 & 3.2\\
\hline\\                
\end{tabular} 
 \begin{flushleft}{
 \textit{Note.} 
In the last column the maximum polarization degree (PD$_{\rm max}$) resulting from the simulations is reported.
}\end{flushleft} 
\end{table}

\begin{figure*}
\includegraphics[width=2.2\columnwidth] {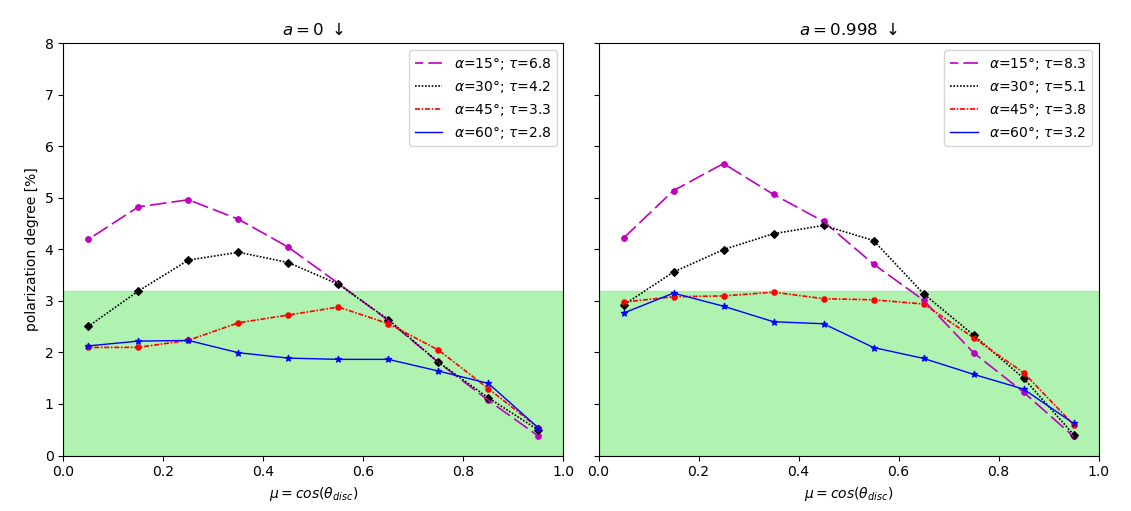}
\caption{Polarization degree from the \textsc{monk} simulations in the case of a wedge-shaped corona as a function of the cosine of the inclination angle ($\mu=\cos\theta_{\rm disc}$, where $\mu=0$ and $\mu=1$ represent the edge-on and face-on views of the source, respectively. \textit{Left panel}: static SMBH ($a=0$) cases. \textit{Right panel}: maximally spinning SMBH ($a=0.998$).  Purple, black, red and blue lines correspond to the 15$\degr$, 30$\degr$, 45$\degr$ and 60$\degr$ opening angles cases, respectively). The green regions represent the allowed values of the polarization degree (see  Sect.\ref{combined_analysis}).}
\label{poldeg}
\end{figure*}

The polarization angle is found to be always parallel to the accretion disc axis. The degree of polarization is up to 5--6 per cent for the smaller opening angles, showing no significant variations with energy in the 2--8 keV energy range. In all tested cases, we notice a decrease in the degree of polarization for larger opening angles, as the geometry becomes closer to a sphere, for which zero polarization is expected. Finally, we notice a slight increase in PD between the static and the maximum spinning black hole cases. In Fig. \ref{poldeg} we show the polarization degree as a function of the cosine of the inclination angle ($\mu=\cos\theta_{\rm disc}$).

\begin{figure*}
\includegraphics[width=1.9\columnwidth] {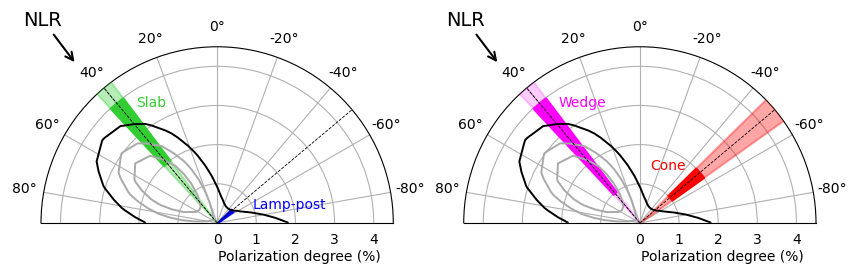}
\caption{Comparison between \textsc{monk} simulations and the contour plot of the combined analysis presented in Sect. \ref{combined_analysis}. Different coronal geometries are shown: slab (in light green) and spherical lamppost (in blue) in the \textit{left panel}, wedge (in magenta) and cone (in red) in the \textit{right panel}. 
Regions of the plot filled with pale colours represent the expected $\Pi$ for all the possible inclinations of the source, while the saturated ones represent the expected degree for inclinations in the range $30\degr-50\degr$, as found in Serafinelli et al. (in prep.).
The black-dotted line at $40\degr$ represents the supposed elongation of the NLR (which is the expected polarization angle in the slab and wedge geometries), while the black-dotted line at $-50\degr$ represents the direction orthogonal to the NLR, (the expected polarization angle for the lamppost and the cone).}\label{geometries}
\end{figure*}

\section{Conclusions}
\label{discussion}

Constraining the geometry of the comptonizing corona in AGNs is one of the main goals of \textit{IXPE}. So far, three    radio-quiet, unobscured AGNs have been observed: MCG-05-23-16 \citep{Marinucci_2022}, NGC\,4151 \citep{gia2023arXiv230312541G} and IC\,4329A (Ingram et al. in prep.).
The first observation of MCG-05-23-16 put constraints on the polarization degree of the primary continuum ($\Pi\le4.7$ per cent) and found a hint of alignment between the polarization angle and the accretion disc spin axis. 
In NGC\,4151 a clear detection has been obtained, with a polarization degree of $\Pi=4.9\pm1.1$ per cent and a polarization angle parallel to the disc axis (as probed by the radio jet). These results disfavour a lamppost geometry  \citep{gia2023arXiv230312541G}

In this paper we have analysed the second pointing of MCG-05-23-16 performed by \textit{IXPE} on November 2022, also combining this observation with the first one (May 2022), and using \textit{XMM-Newton} and \textit{NuSTAR} data taken contemporaneously.
The results were then compared with theoretical simulations performed with the Comptonization Monte Carlo code \textsc{monk}.
The combined analysis led to a significant decrease of the upper limit to the polarization degree of the primary continuum, which is now  $\Pi\le3.2$ per cent (to be compared with  $\Pi\le4.7$ per cent from the first observation only,  \citealt{Marinucci_2022}).

\textit{Hubble Space Telescope}'s WFPC2 images showed that the ionization cone of  MCG-05-23-16 has a roughly 40\degr\ position angle, as probed by [O \textsc{iii}] emission \citep{hub2000ApJS..128..139F}. Let us assume it as a marker for the Narrow Line Region (NLR), and that the NLR elongation axis is perpendicular to the accretion disc. Even if the polarization angle is formally unconstrained, given that we do not have a firm polarization detection, our analysis nevertheless suggests a statistical preference for a polarization angle in the $\sim$50\degr\ direction (see Fig.~\ref{fine_mondo}). This is a hint that the polarization of the primary emission is aligned with the NLR and so parallel to the accretion disc axis, similar to what was found in NGC\,4151 \citep{gia2023arXiv230312541G}. 

Let us now use the PD--PA contour plots to put constraints on the geometrical parameters of the corona. 
In Fig. \ref{geometries} we plot, superimposed to the contour plots, the polarization degree and angle from \textsc{monk} simulations for four different geometries.
The results for the lamppost, cone and slab are taken from \citet{2022MNRAS.510.3674U} and all assume a static black hole, a coronal temperature of 25 keV and the optical depth which best reproduces the observed MCG-05-23-16 spectrum analyzed by \citet{Marinucci_2022}. In the absence of any independent constraint on the source inclination, we cannot formally exclude any geometry, as, for low enough angles, any of them can reproduce a polarization degree close to zero. For the slab and the wedge cases (which have polarization angles parallel to the disc axis), the effective upper limit is 3.2 per cent, and we can constrain the source inclination to be lower than $40\degr$ assuming the slab geometry. If we instead assume the wedge geometry, the allowed range of source inclinations depends also on opening angle $\alpha$. We see from Fig. \ref{poldeg} that for $\alpha \gtrsim 45\degr$, the predicted polarisation degree is always below our observational upper limit, thus leaving the source inclination unconstrained.
On the other hand, assuming a static SMBH, constrains the inclination to be either below $50\degr$ or above $80\degr$ for $\alpha=30 \degr$ and to be lower than $50\degr$ for $\alpha = 15\degr$. Assuming instead a maximally spinning SMBH, constrains the inclination to be lower than about $40\degr$ for both $\alpha = 30\degr$ and $\alpha = 15\degr$.

For coronal geometries that predict polarization angles perpendicular to the disc axis, the upper limit on polarization degree is much more stringent ($\Pi\le0.5$ per cent). In this scenario, if we consider the cone-shaped corona, we can constrain the source inclination to be lower than $20\degr$. Finally, considering the lamppost geometry, since it predicts a very low PD for all inclinations, no constraints could be obtained. 
Information on the inclination angle, however, can in principle be obtained by modeling the reflection component. Serafinelli at al. (in prep.) found the inclination of MCG-05-23-16 to be constrained in the $30\degr-50\degr$ range. If we assume these values, Fig. \ref{geometries} shows that the cone-shaped corona is disfavoured.

\section*{Acknowledgements}
The {\it Imaging X ray Polarimetry Explorer} ({\it IXPE}) is a joint US and Italian mission.  The US contribution is supported by the National Aeronautics and Space Administration (NASA) and led and managed by its Marshall Space Flight Center (MSFC), with industry partner Ball Aerospace (contract NNM15AA18C).  The Italian contribution is supported by the Italian Space Agency (ASI) through contract ASI-OHBI-2017-12-I.0, agreements ASI-INAF-2017-12-H0 and ASI-INFN-2017.13-H0, and its Space Science Data Center (SSDC), and by the Istituto Nazionale di Astrofisica (INAF) and the Istituto Nazionale di Fisica Nucleare (INFN) in Italy.  This research used data products provided by the {\it IXPE} Team (MSFC, SSDC, INAF, and INFN) and distributed with additional software tools by the High-Energy Astrophysics Science Archive Research Center (HEASARC), at NASA Goddard Space Flight Center (GSFC). Part of the French contribution is supported by the Scientific Research National Center (CNRS) and the French Space Agency (CNES). MD, VK and JPod thank for the support from the GACR project 21-06825X and the institutional support from RVO:67985815. I.A. acknowledges financial support from the Spanish "Ministerio de Ciencia e Innovaci\'on” (MCINN) through the “Center of Excellence Severo Ochoa” award for the Instituto de Astrof\'isica de Andaluc\'ia-CSIC (SEV-2017-0709) and through grants AYA2016-80889-P and PID2019-107847RB-C44.

\section*{Data Availability}
The data analyzed in this work are either publicly available at the {\sc HEASARC} database or available from the corresponding author upon request.

\bibliographystyle{mnras}
\bibliography{biblio}

\appendix

\section*{}
\newcounter{foo}  
\begin{list}{$^{\arabic{foo}}$}
    {\usecounter{foo}
     \setlength{\labelwidth}{0em}
     \setlength{\labelsep}{0em}
     \setlength{\itemsep}{0pt}
     \setlength{\leftmargin}{0cm}
     \setlength{\rightmargin}{0cm}
     \setlength{\itemindent}{0em} 
    }
\itshape
\small
\item Dipartimento di Matematica e Fisica, Universit\`a degli Studi Roma Tre, via della Vasca Navale 84, 00146 Roma, Italy  \label{in:UniRoma3}
\item Agenzia Spaziale Italiana, Via del Politecnico snc, 00133 Roma, Italy \label{in:ASI}
\item Department of Physics and Kavli Institute for Particle Astrophysics and Cosmology, Stanford University, Stanford, California 94305, USA  \label{in:Stanford}
\item Istituto Nazionale di Fisica Nucleare, Sezione di Pisa, Largo B. Pontecorvo 3, 56127 Pisa, Italy \label{in:INFN-PI}
\item Dipartimento di Fisica, Universit\`{a} di Pisa, Largo B. Pontecorvo 3, 56127 Pisa, Italy \label{in:UniPI}
\item Universit\'{e} de Strasbourg, CNRS, Observatoire Astronomique de Strasbourg, UMR 7550, 67000 Strasbourg, France \label{in:Strasbourg}
\item Physics Department and McDonnell Center for the Space Sciences, Washington University in St. Louis, St. Louis, MO 63130, USA \label{in:WUStL}
\item INAF Istituto di Astrofisica e Planetologia Spaziali, Via del Fosso del Cavaliere 100, 00133 Roma, Italy \label{in:INAF-IAPS}
\item School of Mathematics, Statistics, and Physics, Newcastle University, Newcastle upon Tyne NE1 7RU, UK \label{in:Newcastle}
\item Astronomical Institute of the Czech Academy of Sciences, Boční II 1401/1, 14100 Praha 4, Czech Republic \label{in:CAS-ASU}
\item Dipartimento di Fisica, Università degli Studi di Roma "La Sapienza," Piazzale Aldo Moro 5, I-00185 Roma, Italy \label{in:SAPIENZA}
\item Dipartimento di Fisica, Universit\`{a} degli Studi di Roma ``Tor Vergata'', Via della Ricerca Scientifica 1, 00133 Roma, Italy \label{in:UniRoma2}
\item INAF Osservatorio Astronomico di Roma, Via Frascati 33, 00040 Monte Porzio Catone (RM), Italy \label{in:INAF-OAR} 
\item Space Science Data Center, Agenzia Spaziale Italiana, Via del Politecnico snc, 00133 Roma, Italy \label{in:ASI-SSDC}
\item MIT Kavli Institute for Astrophysics and Space Research, Massachusetts Institute of Technology, 77 Massachusetts Avenue, Cambridge, MA 02139, USA \label{in:MIT}
\item Universit\'{e} Grenoble Alpes, CNRS, IPAG, 38000 Grenoble, France \label{in:Grenoble}
\item Astronomical Institute, Charles University, V Holešovičkách 2, CZ-18000 Prague, Czech Republic \label{in:Charles} 
\item Department of Physics and Astronomy, FI-20014 University of Turku,  Finland \label{in:UTU} 
\item Istituto Nazionale di Fisica Nucleare, Sezione di Roma ``Tor Vergata'', Via della Ricerca Scientifica 1, 00133 Roma, Italy \label{in:INFN-Roma2}
\item Department of Astronomy, University of Maryland, College Park, Maryland 20742, USA \label{in:UMd} 
\item Nordita, KTH Royal Institute of Technology and Stockholm University, Hannes Alfv\'{e}ns v\"{a}g 12, SE-10691 Stockholm, Sweden \label{in:KTH}
\item National Astronomical Observatories, Chinese Academy of Sciences, 20A Datun Road, Beijing 100101, China \label{in:NAO-CAS}
\item Instituto de Astrof\'{i}sicade Andaluc\'{i}a -- CSIC, Glorieta de la Astronom\'{i}a s/n, 18008 Granada, Spain \label{in:CSIC-IAA}
\item INAF Osservatorio Astronomico di Cagliari, Via della Scienza 5, 09047 Selargius (CA), Italy  \label{in:INAF-OAC} 
\item NASA Marshall Space Flight Center, Huntsville, AL 35812, USA \label{in:NASA-MSFC}
\item Istituto Nazionale di Fisica Nucleare, Sezione di Torino, Via Pietro Giuria 1, 10125 Torino, Italy  \label{in:INFN-TO}	
\item Dipartimento di Fisica, Universit\`{a} degli Studi di Torino, Via Pietro Giuria 1, 10125 Torino, Italy \label{in:UniTO} 
\item INAF Osservatorio Astrofisico di Arcetri, Largo Enrico Fermi 5, 50125 Firenze, Italy \label{in:INAF-Arcetri} 
\item Dipartimento di Fisica e Astronomia, Universit\`{a} degli Studi di Firenze, Via Sansone 1, 50019 Sesto Fiorentino (FI), Italy \label{in:UniFI} 
\item Istituto Nazionale di Fisica Nucleare, Sezione di Firenze, Via Sansone 1, 50019 Sesto Fiorentino (FI), Italy \label{in:INFN-FI}
\item Science and Technology Institute, Universities Space Research Association, Huntsville, AL 35805, USA \label{in:USRA-MSFC}
\item Institut f\"ur Astronomie und Astrophysik, Universit\"at T\"ubingen, Sand 1, D-72076 T\"ubingen, Germany \label{in:Tub}
\item RIKEN Cluster for Pioneering Research, 2-1 Hirosawa, Wako, Saitama 351-0198, Japan \label{in:RIKEN}
\item California Institute of Technology, Pasadena, CA 91125, USA \label{in:Caltech}
\item Yamagata University,1-4-12 Kojirakawa-machi, Yamagata-shi 990-8560, Japan \label{in:Yamagata}
\item University of British Columbia, Vancouver, BC V6T 1Z4, Canada \label{in:UBC}
\item International Center for Hadron Astrophysics, Chiba University, Chiba 263-8522, Japan \label{in:Chiba}
\item Institute for Astrophysical Research, Boston University, 725 Commonwealth Avenue, Boston, MA 02215, USA \label{in:BU} 
\item Department of Astrophysics, St. Petersburg State University, Universitetsky pr. 28, Petrodvoretz, 198504 St. Petersburg, Russia \label{in:SPBU} 
\item Department of Physics and Astronomy and Space Science Center, University of New Hampshire, Durham, NH 03824, USA \label{in:UNH} 
\item Finnish Centre for Astronomy with ESO, 20014 University of Turku, Turku 20014, Finland \label{in:FINCA}
\item Istituto Nazionale di Fisica Nucleare, Sezione di Napoli, Strada Comunale Cinthia, 80126 Napoli, Italy \label{INFN-NA}
\item Graduate School of Science, Division of Particle and Astrophysical Science, Nagoya University, Furo-cho, Chikusa-ku, Nagoya, Aichi 464-8602, Japan \label{in:Nagoya}
\item Hiroshima Astrophysical Science Center, Hiroshima University, 1-3-1 Kagamiyama, Higashi-Hiroshima, Hiroshima 739-8526, Japan \label{in:Hiroshima}
\item University of Maryland, Baltimore County, Baltimore, MD 21250, USA \label{in:UMBC}
\item NASA Goddard Space Flight Center, Greenbelt, MD 20771, USA  \label{in:NASA-GSFC}
\item Center for Research and Exploration in Space Science and Technology, NASA/GSFC, Greenbelt, MD 20771, USA  \label{in:CRESST}
\item Department of Physics, University of Hong Kong, Pokfulam, Hong Kong \label{in:HKU}
\item Department of Astronomy and Astrophysics, Pennsylvania State University, University Park, PA 16801, USA \label{in:PSU}
\item Center for Astrophysics, Harvard \& Smithsonian, 60 Garden St, Cambridge, MA 02138, USA \label{in:CfA} 
\item INAF Osservatorio Astronomico di Brera, via E. Bianchi 46, 23807 Merate (LC), Italy \label{in:INAF-OAB}
\item Dipartimento di Fisica e Astronomia, Universit\`{a} degli Studi di Padova, Via Marzolo 8, 35131 Padova, Italy \label{in:UniPD}
\item Mullard Space Science Laboratory, University College London, Holmbury St Mary, Dorking, Surrey RH5 6NT, UK \label{in:MSSL}
\item Anton Pannekoek Institute for Astronomy \& GRAPPA, University of Amsterdam, Science Park 904, 1098 XH Amsterdam, The Netherlands  \label{in:Amsterdam}
\item Guangxi Key Laboratory for Relativistic Astrophysics, School of Physical Science and Technology, Guangxi University, Nanning 530004, China \label{in:GSU}
\end{list}

\bsp	
\label{lastpage}

\end{document}